\documentclass[journal]{IEEEtran}
\usepackage{amsmath,epsfig}
\usepackage{amsmath,amssymb,amsfonts}
\usepackage{makecell}
\usepackage{url}
\usepackage{graphicx}
\usepackage{textcomp}
\usepackage{cite}
\usepackage{times}
\usepackage{subfigure}
\usepackage{booktabs}
\usepackage{multirow}
\usepackage{caption}
\usepackage{color}
\newcommand{\red}[1]{\textcolor{blue}{#1}}

\usepackage[ruled]{algorithm2e}

\begin{document}
\title{SAR Ship Target Recognition via Selective Feature Discrimination and Multifeature Center Classifier}
%
%
%

\author{Chenwei Wang,~\IEEEmembership{Student Member,~IEEE,}
        Siyi Luo, 
        Jifang Pei,~\IEEEmembership{Member,~IEEE,} \\
        Yulin Huang,~\IEEEmembership{Senior Member,~IEEE,}
        Yin Zhang,~\IEEEmembership{Member,~IEEE,}
        Jianyu Yang,~\IEEEmembership{Member,~IEEE}
\thanks{This work was supported by the National Natural Science Foundation of China under Grants 61901091 and 61901090. (\emph{Corresponding author: Jifang Pei.})

       The authors are with the Department of Electrical Engineering, University of Electronic Science and Technology of China, Chengdu 611731, China (e-mail: peijfstudy@126.com; dbw181101@163.com).}}

\maketitle

\begin{abstract}
Maritime surveillance is not only necessary for every country, such as in maritime safeguarding and fishing controls, but also plays an essential role in international fields, such as in rescue support and illegal immigration control.
Synthetic aperture radar (SAR) ship recognition is an indispensable foundation stone for  maritime surveillance. 
Most of the existing automatic target recognition (ATR) methods directly send the extracted whole features of SAR ships into one classifier. The classifiers of most methods only assign one feature center to each class.
However, the characteristics of SAR ship images, large inner-class variance, and small interclass difference lead to the whole features containing useless partial features and a single feature center for each class in the classifier failing with large inner-class variance.
Thus, the performances of the existing methods are limited. 
This paper proposes a SAR ship target recognition method via selective feature discrimination and multifeature center classifier. 
The selective feature discrimination automatically finds the similar partial features from the most similar interclass image pairs and the dissimilar partial features from the most dissimilar inner-class image pairs.
It then provides a loss to enhance these partial features with more interclass separability. 
Motivated by divide and conquer, the multifeature center classifier assigns multiple learnable feature centers for each ship class. In this way, the multifeature centers divide the large inner-class variance into several smaller variances and conquered by combining all feature centers of one ship class.
In this way, the multifeature center classifier can encompass the large inner-class variance of SAR ship images. 
Finally, the probability distribution over all feature centers is considered comprehensively to achieve an accurate recognition of SAR ship images. 
The ablation experiments and experimental results on OpenSARShip and FUSAR-Ship datasets show that our method has achieved superior recognition performance under decreasing training SAR ship samples.
\end{abstract}

\begin{IEEEkeywords}
synthetic aperture radar (SAR), ship automatic target recognition (ATR), selective feature discrimination, multifeature center classifier
\end{IEEEkeywords}

%
\IEEEpeerreviewmaketitle

\section{Introduction}
\IEEEPARstart{M}{aritime} surveillance is an indispensable and crucial mission for both military and civilian fields, including maritime disaster surveillance, channel monitoring, and national maritime safeguarding, among others \cite{intro1ms}. 
Ship monitoring is the foundation stone for many maritime surveillance tasks and attracts more attention from researchers \cite{intro1,wang2022recognition,wang2023entropy,intro6,intro7,intro8}. The current transponder-based ship monitoring systems made up of automatic identification systems (AIS) and vessel traffic services (VTS) are inevitably problematic in some unexpected or uncooperative situations. In recent years, synthetic aperture radar (SAR) has become one of the effective means of ship monitoring as it can provide high-resolution, weather-independent images throughout the day \cite{intro24}. Therefore, researches about SAR ship monitoring that which mainly focus on ship detection and recognition are necessary and critical \cite{intro3,wang2023sar,intro4}. In contrast to the vast research in ship detection, SAR ship recognition is still an important issue that requires more attention from the ship monitoring community \cite{intro23,wang2022global,wang2022semi,intro22,intro5}.

\begin{figure*}[thb]
\centering
\subfigbottomskip=0pt
\subfigcapskip=-3pt
\subfigure[]{
    \label{intro1:a}
    \includegraphics[width=0.32\linewidth]{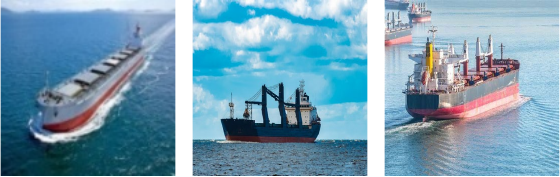} 
    } 
\subfigure[]{
    \label{intro1:b}
    \includegraphics[width=0.64\linewidth]{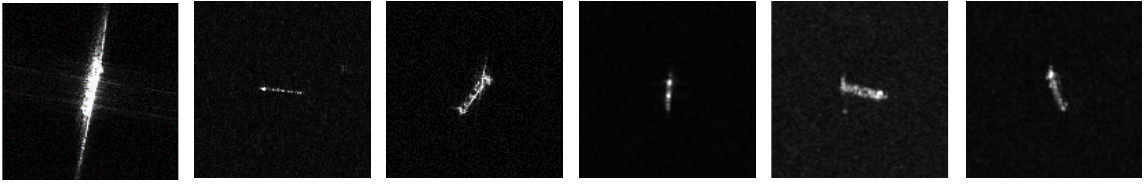} 
    }\\
\subfigure[]{
    \label{intro1:c}
    \includegraphics[width=0.32\linewidth]{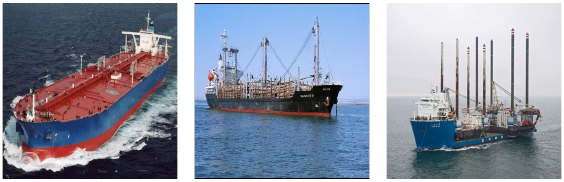} 
    }
\subfigure[]{
    \label{intro1:d}
    \includegraphics[width=0.64\linewidth]{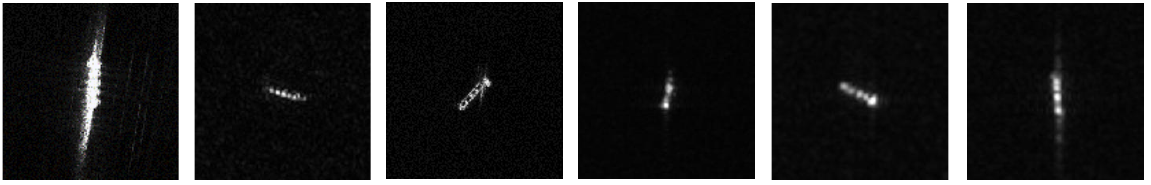} 
    }\\
\subfigure[]{
    \label{intro1:e}
    \includegraphics[width=0.32\linewidth]{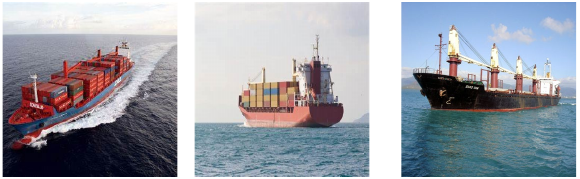} 
    }
\subfigure[]{
    \label{intro1:f}
    \includegraphics[width=0.64\linewidth]{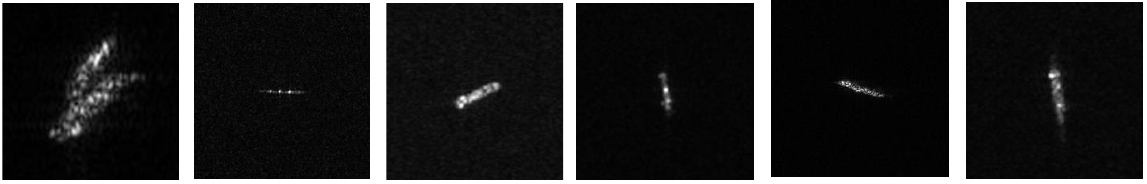} 
    }
\bfseries\caption{Images of bulk carriers, general cargoes, and container ships. (a) and (b) show optical and SAR images of bulk carriers; (c) and (d) show optical and SAR images of general cargoes; (e) and (f) show optical and SAR images of container ships.}
\label{intro1}
\end{figure*}

The problems of SAR ship target recognition can be summarized as the large inner-class variance and small interclass differences, by analyzing from two aspects, i.e., actual ship variety and SAR ship imaging condition \cite{intro2,intro20,open2,open3}. (1) From the actual ship variety, the inner-class ships may have a large variance in size and appearance, and the interclass ships have a large overlap in size and appearance. For example, the size of general cargoes can widely range from lengths of 90-200m and widths of 15-33m, but the size of bulk carriers ranges from lengths of 150-275m and widths of 23-38m. The appearances of two general cargoes have an obvious difference, as shown in Fig. \ref{intro1}(a); however, the ships of general cargoes and bulk carriers have a very similar appearance, as shown by comparing Fig. \ref{intro1}(a) and (b).
(2) From the SAR ship imaging condition, the changing imaging conditions, such as resolution and incident angle, make the image feature of inner-class SAR ship images vary, as shown in Fig. \ref{intro1}(c) and (d). At the same time, the crucial image features for discriminating different ship classes may be occluded in some ranges of depression angle or azimuth, which also increase the difficulty of recognizing different ship classes.  

The large inner-class variance and small interclass differences of SAR ship images impose two requirements on the recognition model: (1) the model can correctly recognize the same ship classes from the inner-class features with large variance, and (2) the model can correctly recognize different ship classes from the interclass features with small differences.  

In recent years, some pioneering researches are conducted to tackle the problem of SAR ship target recognition \cite{intro9,wang2020deep,intro10,wang2022sar,wang2021multiview,wang2019parking,intro11,wang2021deep,intro12}. For example, Tirandaz and Akbarizadeh \cite{intro13} proposed a two-stage texture-based approach that can be used for SAR ship target recognition. Considering the information content of medium-resolution SAR images and the stability of feature extraction, Lang and Wu \cite{intro14} proposed naive geometric features (NGFs) for ship recognition. In addition, Lin \textit{et al.} \cite{intro15} designed a feature called manifold-learning SAR histogram of oriented gradients (MSHOG) by improving the SAR histogram of oriented gradients (SAR-HOG). Based on this, they proposed a ship classification method, i.e., MSHOG feature and task-driven dictionary learning (TDDL) with structured incoherent constraints.  

However, it is hard for these handcrafted extracted features to be always robust and effective, leading to poor generalization of these methods. In light of the fast development of deep learning, some researches are also conducted focusing on the problem of SAR ship target recognition \cite{intro16,wang2020multi,li2023panoptic,intro17,liang2023efficient,intro18}. For example, Li \textit{et al.} \cite{inrto19} designed a dense residual network with several optimized training techniques for the limitation of the unbalanced dataset. He \textit{et al.} \cite{intro20} extended the dense convolutional networks and proposed a multitask learning framework for more effective ship recognition in medium-resolution SAR images. Xu and Lang \cite{intro21} proposed a transfer metric learning method, i.e., geometric transfer metric learning (GTML), to improve SAR ship target recognition performance.  

Most of existing deep learning-based methods proposed novel technologies and achieved good SAR ship target recognition performance. These methods input the extracted whole features of SAR ship images into the classifier to complete the recognition. However, as mentioned above, the small interclass differences and large inner-class variance of SAR ship images lead to the following: (1) the interclass SAR ship images may have partially similar features, which interferes with accurate recognition, and (2) the common classifier, such as SoftMax, which actually assumes one single center for each class \cite{method1}, will not be effective for the features with large inner-class variance. Therefore, the performances of these methods are limited.  

Therefore, this paper proposes a SAR ship target recognition method via selective feature discrimination and multifeature center classifier to tackle the problems mentioned above. The designed method can be illustrated as follows, as shown in Fig. \ref{framework}. 

\noindent 1. First, the proposed method perceives the whole SAR ship images and gradually extracts and integrates local image features layer by layer to obtain the hierarchical ship features. 
The selective feature discrimination then automatically acquires partial features that are similar to other ship types among all features of every SAR ship image and enhances these features for every SAR ship image with more interclass separability.

\noindent 2. The multifeature center classifier first sets multiple different learnable feature centers for each ship type and then makes the inner-class features with a large variance close to one of the feature centers of that class in the training process, so that the multifeature centers divide the large variance into several smaller variances. In this way, these inner-class feature centers can encompass the large inner-class variance of SAR ship images. Finally, the probability distribution of SAR ship features over all feature centers is calculated and considered comprehensively to complete the recognition.

The main contributions of this paper are summarized as follows:

(1) The selective feature discrimination enhances features with more interclass separability by enlarging the distance of similar partial features from most similar image pairs of different ship classes.

(2) The multifeature center classifier sets multiple learnable feature centers for each ship type to encompass large inner-class feature variations, thus accomplishing accurate recognition.

(3) The proposed method achieves the state-of-the-art performance of recognition on OpenSARShip and FUSAR-Ship datasets which have the problem mentioned above. Furthermore, the robustness of the method is also validated under decreasing training  samples. The ablation experiments and the comparisons with other ship recognition methods and famous deep learning methods validate the effectiveness of our methods.


The remainder of this paper is organized as follows. Section \uppercase\expandafter{\romannumeral2} presents an overview of the proposed method.  The experiments in Section \uppercase\expandafter{\romannumeral3} validate the effectiveness of the proposed method. Finally, Section \uppercase\expandafter{\romannumeral4} provides a conclusion.

\section{Proposed Method}
This section describes the framework of the proposed method in detail and also presents the selective feature discrimination. The multifeature center classifier is then described in particular.

\subsection{Framework of Proposed Method(PLEASE BE NOTED THAT THE PROPOSED METHOD HAS BEEN REVISED IN THE NEWEST SUBMITTED VERSION )}
As analyzed above, the large inner-class variance and small interclass differences of SAR ship images lead to two subproblems: (1) partial features are useless, and (2) a single feature center is not suitable for SAR ship images. The selective feature discrimination and the multifeature center classifier are proposed for the first and second subproblems, respectively.


As shown in Fig. \ref{framework}, the whole framework of the proposed method consists of three parts: extractor, selective feature discrimination, and multifeature center classifier. 
The extractor is designed to construct the hierarchical embedding, widen the receive fields, and extract the optimal feature maps to provide robust and effective features for the subsequent modules and the final recognition.

The selective feature discrimination aims to enhance the interclass similar partial features from different ship classes, as shown in orange box in Fig. \ref{framework}. Firstly, it finds the most similar interclass image pairs and the most dissimilar inner-class image pairs. It then automatically finds the similar partial features from this interclass image pairs and enlarges the distances of these partial features and vice versa. Through the selective feature discrimination, the useless partial features are found and enhanced with more discriminativeness. In this way, the whole features of a SAR ship image can be enhanced with more inner-class compactness and interclass separability.

The multifeature center classifier aims to tackle the inner-class features with large variance, as shown in the blue box in Fig. \ref{framework}. Firstly, it sets multiple different learnable feature centers for each ship class. In the training process, the inner-class features with large variance are then assigned to one of the feature centers, so that the multifeature centers divide the large variance into several smaller variances. The multifeature center classifier then integrates the probability distribution over these feature centers together and finishes the final recognition, so that it can deal with the problem of the large variance and small compactness from the same ship classes.

The mathematical representation of the proposed method is described in detail as follows. Given a SAR ship image $\mathbf{x}_i^l$, where $\mathbf{x}_i^l$ is the $l$th SAR ship image from the $i$th ship class, by inputting $\mathbf{x}_i^l$ into the extractor, the output feature maps can be presented as $\mathbf{M}\left( \mathbf{x}_{i}^{l} \right)\in {{\mathbb{R}}^{h\times w\times c}}$, where $h$, $w$, and $c$ are the length, width, and channel number of the feature maps, respectively.

First, the selective feature discrimination ranks the similarities between $\mathbf{M}\left( \mathbf{x}_{i}^{l} \right)$ and the feature maps of the other SAR ship classes $\mathbf{M}\left( \mathbf{x}_{j}^{l} \right)$ to find the most similar interclass feature maps and the most dissimilar feature maps $\mathbf{M}\left( \mathbf{x}_{i}^{k} \right)$ of the $i$th ship class. 

Second, the selective feature discrimination finds the similar partial features between $\mathbf{M}\left( \mathbf{x}_{i}^{l} \right)$ and $\mathbf{M}\left( \mathbf{x}_{j}^{l} \right)$ and the dissimilar partial features between $\mathbf{M}\left( \mathbf{x}_{i}^{l} \right)$ and $\mathbf{M}\left( \mathbf{x}_{i}^{k} \right)$. Finally, the discrimination provides a loss $L_{disc}$ to enhance the inner-class compactness and interclass separability of the partial features.

Afterward, the feature maps $\mathbf{M}\left( \mathbf{x}_{i}^{l} \right)$ go through the global average pooling (GAP) layer \cite{GAP} and dense layer and become the feature vectors $\mathbf{V}\left( \mathbf{x}_{i}^{l} \right)\in {{\mathbb{R}}^{1\times 1\times c}}$. The classifier firstly normalizes $\mathbf{V}\left( \mathbf{x}_{i}^{l} \right)$, like L2 normalization, and calculates the distance between $\mathbf{V}\left( \mathbf{x}_{i}^{l} \right)$ and the feature centers with scaling to provide a loss ${L}_{mfc}$ to finish the final recognition. The total loss can be calculated as follows:
\begin{equation}
L={{\lambda }_{1}}{{L}_{mfc}}+{{\lambda }_{2}}{{L}_{disc}}
\end{equation}
where ${\lambda_1}$ and ${\lambda_2}$ are the weighting coefficients. ${{L}_{disc}}$ enhances the partial features which are \red{useless} for recognition, and ${{L}_{mfc}}$ optimizes the inner-class feature centers of each ship class to encompass the large inner-class variance.

Through the process, the proposed method can tackle the large inner-class variance and small interclass difference to achieve an accurate recognition of SAR ship images.
The details of the selective feature discrimination and the multifeature center classifier are described as follows.

\subsection{Selective Feature Discrimination}
The useless partial features limit the discrimination of the whole features of SAR ship images, further affecting the recognition performance.
Therefore, the selective feature discrimination aims to firstly find the partial features that contribute minimally to recognition or interfere with it, among all the features of the SAR ship images. It then enhances the discriminativeness of these partial features, thereby improving the recognition performance of SAR ship images.

Given the SAR ship images $\left\{ {\bf{x}}_{1}^{1},{\bf{x}}_{1}^{2},\cdots ,{\bf{x}}_{n}^{k} \right\}$, ${\bf{x}}_{i}^{j}$ means the $j\text{th}$ SAR ship images of the $i\text{th}$ ship class, and the corresponding features $f({\bf{x}}_{i}^{j})\in {{\mathbb{R}}^{n\times c}}$ extracted from ${\bf{x}}_{i}^{j}$. The construction process of ${{L}_{disc}}$ mainly consists of three steps: (1) finding the most similar interclass image pairs $\left( {\bf{x}}_{i}^{l},{\bf{x}}_{i}^{k} \right)$ and the most dissimilar inner-class image pairs $\left( {\bf{x}}_{i}^{l},{\bf{x}}_{j}^{m} \right)$, (2) finding the dissimilar/similar partial features of $\left( {\bf{x}}_{i}^{l},{\bf{x}}_{i}^{k} \right)$ or $\left( {\bf{x}}_{i}^{l},{\bf{x}}_{j}^{m} \right)$, and (3) providing the loss ${{L}_{disc}}$ to optimize these partial features for better inner-class compactness and interclass separability. The details of the three steps are as follows.

Step 1: The construction process first finds the most similar interclass image pairs and the most dissimilar inner-class image pairs. This step calculates the cosine similarity of the features extracted from the SAR ship images using the following equation to obtain the confusing inner-class image pair ${{S}_{inner}}$ and interclass image pair ${{S}_{inter}}$:

\begin{equation}
{{S}_{inner}}=\left\{ ({\bf{x}}_{1}^{1},{\bf{x}}_{1}^{o}),\cdots ,({\bf{x}}_{i}^{j},{\bf{x}}_{i}^{m}),\cdots ,({\bf{x}}_{n}^{k},{\bf{x}}_{n}^{e}) \right\}
\end{equation}
\begin{equation}
{{S}_{inter}}=\left\{ ({\bf{x}}_{1}^{1},{\bf{x}}_{p}^{m}),\cdots ,({\bf{x}}_{i}^{j},{\bf{x}}_{l}^{q}),\cdots ,({\bf{x}}_{n}^{k},{\bf{x}}_{r}^{t}) \right\}
\end{equation}
\begin{equation}
sim\left( \mathbf{x}_{i}^{l}\mathbf{,x}_{i}^{k} \right) = \frac{\mathbf{V}\left( \mathbf{x}_{i}^{l} \right)}{{{\left\| \mathbf{V}\left( \mathbf{x}_{i}^{l} \right) \right\|}_{2}}}\cdot \frac{\mathbf{V}\left( \mathbf{x}_{i}^{k} \right)}{{\left\| \mathbf{V}\left( \mathbf{x}_{i}^{k} \right) \right\|}_{2}} 
\end{equation}
where ${{\left\| \cdot  \right\|}_{2}}$ means the $\rm{L}_2$-norm. Then for $\mathbf{x}_{i}^{l}$, the most similar image from different ship classes $\mathbf{x}_{j}^{l}$ and the least similar image in the same ship class $\mathbf{x}_{i}^{k}$ can be found by ranking the similarities.Using cosine similarity in SAR ship recognition offers several advantages, make the proposed method suitable for SAR ship images with varying backscatter feature intensity.

Step 2: The construction process then finds the dissimilar partial features of the dissimilar inner-class SAR ship image pairs and the similar partial features of the similar inter-class SAR ship image pairs. The extracted features $f({\bf{x}}_{i}^{j})=\left\{ {{\bf{v}}^{1}}\left( {\bf{x}}_{i}^{j} \right),\cdots ,{{{\bf{v}}}^{n}}\left( {\bf{x}}_{i}^{j} \right),\cdots ,{{{\bf{v}}}^{k}}\left( {\bf{x}}_{i}^{j} \right) \right\}$, where $f({\bf{x}}_{i}^{j})\in {{\mathbb{R}}^{n\times c}}$ and ${{{\bf{v}}}^{k}}\left( {\bf{x}}_{i}^{j} \right)\in {{\mathbb{R}}^{1\times c}}$. For the feature pairs corresponding to each image pair in ${{S}_{inner}}$ and ${{S}_{inter}}$, such as $\left( f\left( {\bf{x}}_{i}^{j} \right),f\left( {\bf{x}}_{i}^{m} \right) \right)$, the cosine similarity of all ${{{\bf{v}}}^{n}}\left( {\bf{x}}_{i}^{j} \right)$ in $f\left( {\bf{x}}_{i}^{j} \right)$ and all ${{{\bf{v}}}^{n}}\left( {\bf{x}}_{i}^{m} \right)$ in $f\left( {\bf{x}}_{i}^{m} \right)$ is calculated. In this way, the dissimilar feature positions corresponding to each inner-class image pair in ${{S}_{inner}}$ and the similar feature positions corresponding to each interclass image pair in ${{S}_{inter}}$ can be found. For example, ${{\bf{P}}^{pos}}= \left\{ p_{1}^{pos},p_{2}^{pos},\cdots ,p_{{\bf{v}}}^{pos} \right\}$ is found for the inner-class pair $\left( {\bf{x}}_{i}^{l},{\bf{x}}_{i}^{k} \right)$ of ${\bf{x}}_{i}^{l}$, and ${{\bf{P}}^{neg}}=\left\{ p_{1}^{neg},p_{2}^{neg},\cdots ,p_{q}^{neg} \right\}$ is found for the interclass pair $\left( {\bf{x}}_{i}^{l},{\bf{x}}_{j}^{m} \right)$ of ${\bf{x}}_{i}^{l}$. The similarity on the $p\text{th}$ feature maps can be calculated as

\begin{equation}
sim_{partial}^{p}\left( \mathbf{x}_{i}^{l}\mathbf{,x}_{j}^{l} \right) = \frac{f\left( {{\mathbf{M}}^{p}}\left( \mathbf{x}_{i}^{l} \right) \right)}{{{\left\| f\left( {{\mathbf{M}}^{p}}\left( \mathbf{x}_{i}^{l} \right) \right) \right\|}_{2}}} \cdot \frac{f\left( {{\mathbf{M}}^{p}}\left( \mathbf{x}_{j}^{l} \right) \right)}{{{\left\| f\left( {{\mathbf{M}}^{p}}\left( \mathbf{x}_{j}^{l} \right) \right) \right\|}_{2}}} 
\end{equation}
where $f\left(  \cdot  \right)$ is the flattening feature maps and ${{\mathbf{M}}^{p}}\left( \mathbf{x}_{i}^{l} \right)$ is the feature map on the $p\text{th}$ channel. The similar partial feature maps between ${{\mathbf{M}}}\left( \mathbf{x}_{i}^{l} \right)$ and ${{\mathbf{M}}}\left( \mathbf{x}_{j}^{l} \right)$ are obtained by finding the channel set ${{\mathbf{P}}^{neg}}=\left\{ p_{1}^{neg},p_{2}^{neg},\ldots ,p_{q}^{neg} \right\}$ satisfying $sim_{partial}^{p}\left( \mathbf{x}_{i}^{l}\mathbf{,x}_{j}^{l} \right)>0$. Moreover, the dissimilar partial feature maps between $\mathbf{M}\left( \mathbf{x}_{i}^{l} \right)$ and $\mathbf{M}\left( \mathbf{x}_{i}^{k} \right)$ are obtained by finding channel set ${{\mathbf{P}}^{pos}}=\left\{ p_{1}^{pos},p_{2}^{pos},\ldots ,p_{v}^{pos} \right\}$ satisfying $sim_{partial}^{p}\left( \mathbf{x}_{i}^{l}\mathbf{,x}_{j}^{l} \right)<0$.

Step 3: The selective feature discrimination provides the loss ${{L}_{disc}}$ to distance the similar partial features of these interclass image pairs and cluster the dissimilar partial features of these inner-class image pairs, as shown in the equation following.
\begin{equation}
\begin{split}
{{L}_{disc}}\left( {\bf{x}}_{i}^{l} \right) = & \max \left( \frac{1}{q}\sum\limits_{u\in {{\mathbf{P}}^{neg}}}{sim\left( {\mathbf{x}}_{i}^{l},{\mathbf{x}}_{j}^{l} \right)} + \psi - \right.\\
&\phantom{=\;\;}\left. \frac{1}{v}\sum\limits_{u\in {{\mathbf{P}}^{pos}}}{sim\left( \mathbf{x}_{i}^{l},\mathbf{x}_{i}^{k} \right)},0 \right)
\end{split}
\end{equation}
where $q$ and $v$ represent the element numbers of ${{\mathbf{P}}^{neg}}$ and ${{\mathbf{P}}^{pos}}$, respectively, and $\psi $ represents the margin of the feature distances between the different classes and the same classes, which can provide the possibility of accurate recognition.

\begin{figure}[thb]
\centering
\includegraphics[width=0.95\linewidth]{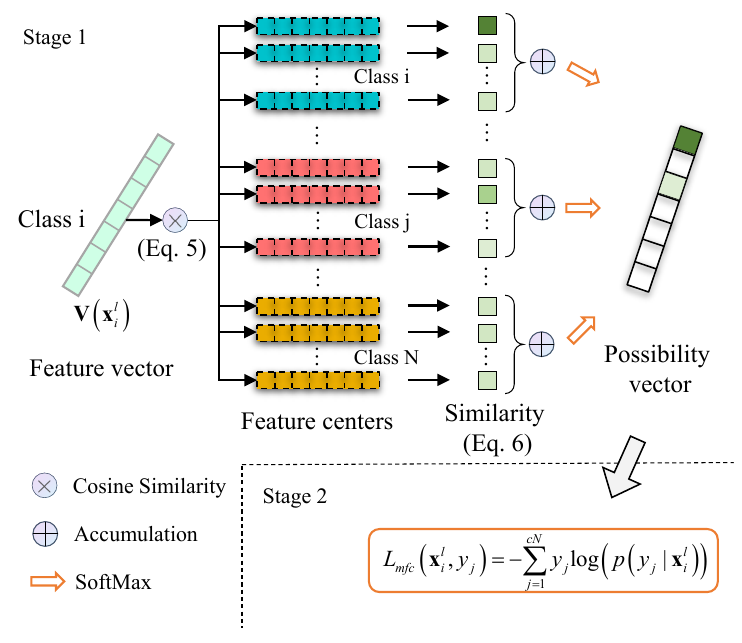} 
\caption{Diagram of the multifeature center classifier. Stage 1 assigns variant features to different feature centers. Stage 2 finishes recognitions by integrating centers. Each square in the probability vector represents a probability value. The darker is the color of the square, the closer the probability value is to 1.}
\label{classifier}
\end{figure}

Therefore, the selective feature discrimination can find the useless partial features that limit the performance of SAR ship image recognition and optimize these partial features during the training process. As a result, the selective feature discrimination can improve the interclass separability of these features for each SAR ship image. 
In this way, it improves the discrimination of the whole features of SAR ship images to achieve an accurate recognition of SAR ship targets. The process and details of the multifeature center classifier will then be presented as follows.

\subsection{Multifeature Center Classifier}
In the real world, the variety of the actual ship size and appearance inevitably lead to the inner-class feature with large variance. Common classifiers, such as SoftMax, which assign the variant inner-class features into only one feature center, will fail to achieve high recognition of SAR ship images.

As shown in the second blue block of Fig. \ref{framework}, the multifeature center classifier provides robust multifeature centers for the inner-class feature with large variance. As shown in Fig. \ref{classifier}, the multifeature center classifier has three stages: initializing multiple centers for each class, assigning features to multiple centers, and finishing recognition by integrating centers.

\begin{figure}[thb]
\centering
\includegraphics[width=0.95\linewidth]{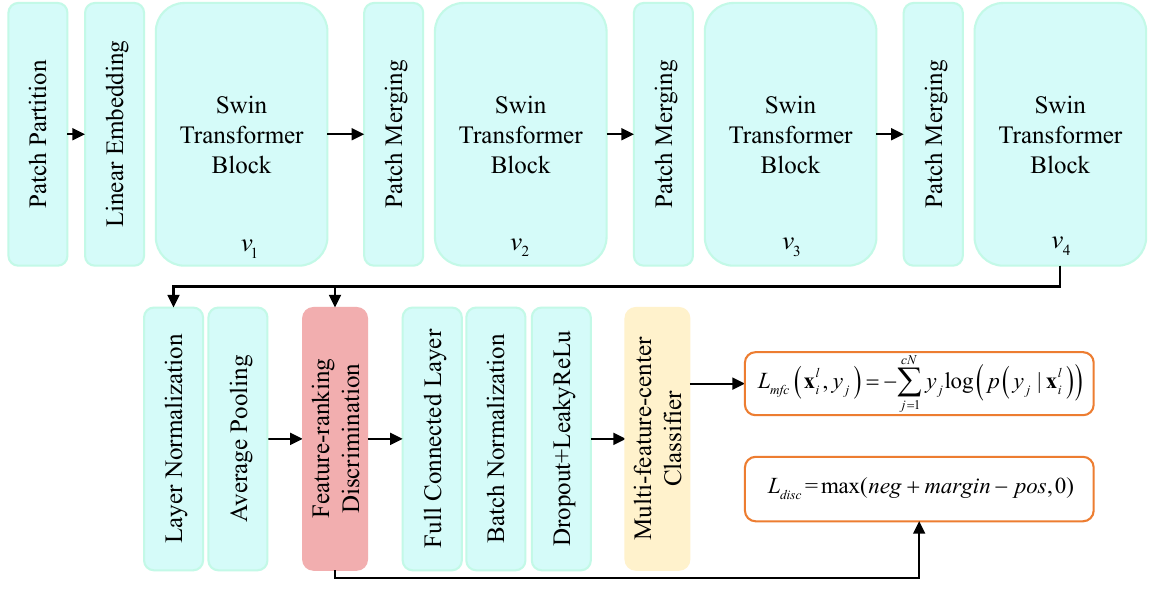} 
\caption{Structure and implementation of ourthe proposed method.}
\label{structure}
\end{figure}

\begin{algorithm}[h]
	\caption{Training and Testing Procedure}
        \label{algortithm_trte}
	\small
     \SetAlgoLined
     \KwIn{SAR ship training set $\mathbf{D}_{train}=\left \{ {\bf{x}}_1^1,\ldots,{\bf{x}}_K^N \right \}$, ${\bf{x}}_K^N$ is $K$th SAR ship image from the $N$th ship class\;
        SAR ship testing set $\mathbf{D}_{test}=\left \{ {\bf{x}}_1^{test},\ldots,{\bf{x}}_Q^{test} \right \}$;}
	\KwOut{predicted ship classes $\mathbf{y}$ of $\mathbf{D}_{test}$\;}
         \BlankLine 
        Initialize: feature extractor $E$, feature centers $\mathbf{C}=\left\{\mathbf{C}_{1}^{1},\ldots,\mathbf{C}_{H}^{N}\right\}$\;
       \nl \While{training}
        {
        \nl sample randomly a batch $\mathbf{B}$ from $D_{train}$, set ${L}_{disc}=0$, ${{L}_{mfc}}=0$\;
        \nl \For{all ${\bf{x}}_i^l$ in $\mathbf{B}$}
        {
        \nl input ${\bf{x}}_i^l$ into $E$\;
        \nl  obtain feature maps $\mathbf{M}\left( \mathbf{x}_{i}^{l} \right)= \left \{{\mathbf{v}^1(\left( \mathbf{x}_{i}^{l} \right)),\ldots,\mathbf{v}^c(\left( \mathbf{x}_{i}^{l} \right))} \right \} \in {\mathbb{R}}^{hw\times c}$, $\mathbf{v}^c(\left( \mathbf{x}_{i}^{l} \right)) \in {\mathbb{R}}^{hw\times 1}$\;
        \textbf{Partial Feature Discrimination:*} \\
        \nl   calculate Eq. 2 between $\mathbf{x}_{i}^{l}$ and other samples in $\mathbf{B}$ to find the most similar inter-class sample ${\bf{x}}_j^o$ and the least similar inner-class sample ${\bf{x}}_t^l$ to $\mathbf{x}_{i}^{l}$\;
        \nl   calculate Eq. 3 between $\mathbf{x}_{i}^{l}$ and ${\bf{x}}_j^o$/${\bf{x}}_t^l$ \;
        \nl   organize ${{\mathbf{P}}(\mathbf{x}_{i}^{l},\mathbf{x}_{j}^{o})=\left\{ p_{1}^n,p_{2}^n,\ldots ,p_{q}^n \right\}}$ by satisfying $sim_{partial}^{p}\left( \mathbf{x}_{i}^{l}\mathbf{,x}_{j}^{o} \right)>0$ in Eq. 3\;
        \nl   organize ${{\mathbf{P}}(\mathbf{x}_{i}^{l},\mathbf{x}_{t}^{l})=\left\{ p_{1}^p,p_{2}^p,\ldots ,p_{v}^p \right\}}$ by satisfying $sim_{partial}^{p}\left( \mathbf{x}_{i}^{l}\mathbf{,x}_{t}^{l} \right)<0$ in Eq. 3\;
        \nl   calculate $L_{disc}(\mathbf{x}_{i}^{l})$ by Eq. 4 \textbf{*} \;
        \textbf{Multi-Feature-Center Classifier:*} \\
        \nl   Calculate the similarity scores between ${\bf{M}}\left( {\bf{x}}_i^l \right )$ and $\mathbf{C}$ by Eq. 5 \;
        \nl   Calculate Eq. 6 to get the possibility of ${\bf{x}}_i^l$ belonging to each ship class\;
        \nl   Calculate Eq. 7 to get ${{L}_{mfc}}(\mathbf{x}_{i}^{l})$\textbf{*}\;
        \nl   calculate ${L}_{disc}+=L_{disc}(\mathbf{x}_{i}^{l})$, ${{L}_{mfc}}+={{L}_{mfc}}(\mathbf{x}_{i}^{l})$\;
        }
        \nl   calculate $L={{\lambda }_{1}}{{L}_{mfc}}+{{\lambda }_{2}}{{L}_{disc}}$\;
        \nl   Update the parameters of $E$ and $\mathbf{C}$ to minimize $L$\;
        }
        \setcounter{AlgoLine}{0}    
        \nl \While{tesing}
        {
            \nl \For{all ${\bf{x}}_j$ in $\mathbf{D}_{test}$}
        {
                \nl input ${\bf{x}}_j$ into $E$\;
                \nl obtain feature maps $\mathbf{M}\left( \mathbf{x}_{j} \right)= \left \{{\mathbf{v}^1(\left( \mathbf{x}_{j} \right)),...,\mathbf{v}^c(\left( \mathbf{x}_{j} \right))} \right \} \in {\mathbb{R}}^{hw*c}$, $\mathbf{v}^c(\left( \mathbf{x}_{j} \right)) \in {\mathbb{R}}^{hw*1}$\;
                \nl   Calculate the similarity scores between ${\bf{M}}\left( {\bf{x}}_j \right )$ and $\mathbf{C}$ by Eq. 5 \;
                \nl   Calculate Eq. 6 to get the possibility of ${\bf{x}}_j$ belonging to each ship class\;
                \nl   obtain the predicted ship class $y({\bf{x}}_j)$\;
        }
                \nl \Return $\mathbf{y}=\left \{y({\bf{x}}_1),...,y({\bf{x}}_Q)\right \}$\\
        }
\end{algorithm}

Stage 1: The Gram-Schmidt orthogonalization method is used to initialize the $H$ orthogonal centers for each class. The steps are presented as follows: (1) Randomly initializing the first centers as the first orthogonal vector $\mathbf{v}_1$; (2) initializing a new vector $\mathbf{v}_2$, projecting it onto $\mathbf{v}_1$ first, and then subtracting the projection of $\mathbf{v}_2$ onto $\mathbf{v}_1$ from $\mathbf{v}_1$ to obtain the second center, denoted as $\mathbf{u}_1$; (3) initializing the third vector $\mathbf{v}_3$ , projecting it onto the plane spanned by  $\mathbf{v}_1$ and $\mathbf{u}_2$, and then subtracting the projection of $\mathbf{v}_3$ onto this plane from $\mathbf{v}_3$ to obtain the third center, denoted as $\mathbf{u}_3$. The above steps are repeated $H$ times to initialize the $H$ orthogonal vectors for each ship class.

Stage 2: The feature maps $\mathbf{M}\left( \mathbf{x}_{i}^{l} \right)$ go through a GAP layer and a dense layer, and become feature vectors $\mathbf{V}\left( \mathbf{x}_{i}^{l} \right)\in {{\mathbb{R}}^{1\times 1\times C}}$. The classifier firstly initializes $H$ learnable feature centers for $N$ ship classes, $\mathbf{C}=\left\{ \mathbf{C}_{1}^{1},\mathbf{C}_{2}^{1},\ldots , \mathbf{C}_{H}^{N} \right\}\in {{\mathbb{R}}^{C\times \Gamma }}$, where $\Gamma $ is the product of the number of ship classes $N$ and $H$. The distance between $\mathbf{V}\left( \mathbf{x}_{i}^{l} \right)$ and all the feature centers is then calculated by

\begin{equation}
simscore\left( \mathbf{x}_{i}^{l}, \mathbf{C}_j^m \right) = \frac{\mathbf{V}\left( \mathbf{x}_{i}^{l} \right)}{{{\left\| \mathbf{V}\left( \mathbf{x}_{i}^{l} \right) \right\|}_{2}}}\cdot \frac{\mathbf{C}_j^m}{{\left\| \mathbf{C}_j^m \right\|}_{2}} 
\end{equation}
where $simscore\left( \mathbf{x}_{i}^{l} \right)\in {{\mathbb{R}}^{1\times \Gamma }}$, $\mathbf{C}_j^m $ is the $j\text{th}$ feature centers of the $m\text{th}$ ship class. Then $simscore\left( \mathbf{x}_{i}^{l},\mathbf{C}_j^m \right)$ is reshaped as ${{\mathbb{R}}^{N\times H}}$, which means the similarity score between the feature vector $\mathbf{V}\left( \mathbf{x}_{i}^{l} \right)$ and the feature centers of all the ship classes. The possibility of $\mathbf{x}_{i}^{l}$ belonging to the $m\text{th}$ ship class can be calculated by integrating the similarity score with SoftMax as

\begin{equation}
p\left( {{y}_{m}}\left| \mathbf{x}_{i}^{l} \right. \right)=\sum\limits_{j=1}^{N}{sf\left( simscor{{e}}\left( \mathbf{x}_{i}^{l}, \mathbf{C}_j^m \right) \right)}-\delta 
\end{equation}
where $sf\left( \cdot  \right)$ is the SoftMax function and $\delta $ is the margin for the robust recognition. When $\mathbf{x}_{i}^{l}$ is classified right, $\delta $ is set as a constant. When $\mathbf{x}_{i}^{l}$ is classified wrong, $\delta $ is set as zero. It should be noted that that $\delta $ can improve the intra-class dispersion and interclass separation of the multiple center of each class in the training process.

Stage 3: The classifier loss is calculated as follows:
\begin{equation}
{{L}_{mfc}}\left( \mathbf{x}_{i}^{l},{{y}_{j}} \right) = -\sum\limits_{j=1}^{N}{{{y}_{j}}}\text{log}\left( p\left( {{y}_{j}}|\mathbf{x}_{i}^{l} \right) \right)
\end{equation}

The proposed method firstly constructs the hierarchical embedding to extract the optimal feature maps.
The selective feature discrimination then enhances useless partial features with more inner-class compactness and interclass separability to tackle the interclass feature with a small difference.
At the same time, the multifeature center classifier assigns learnable feature centers to encompass the large inner-class variance of SAR ship images.
Finally, the proposed method can achieve accurate recognition of SAR ship targets.
Fig. \ref{structure} shows the structure and process of the method in detail to the better understanding and recurrence of the method.
An algorithm chart is organized, as shown in Algorithm \ref{algortithm_trte}, to easily follow the whole procedure. The training and testing processes of the proposed method are shown separately. In detail, Algorithm \ref{algortithm_trte} also presents the two new modules, i.e., selective feature discrimination and multifeature center classifier.

Experiments with the decreasing training data will be conducted in the next section to validate the effectiveness and practicability of the proposed method.

\begin{figure}[tbh]
\centering
\includegraphics[width=0.99\linewidth]{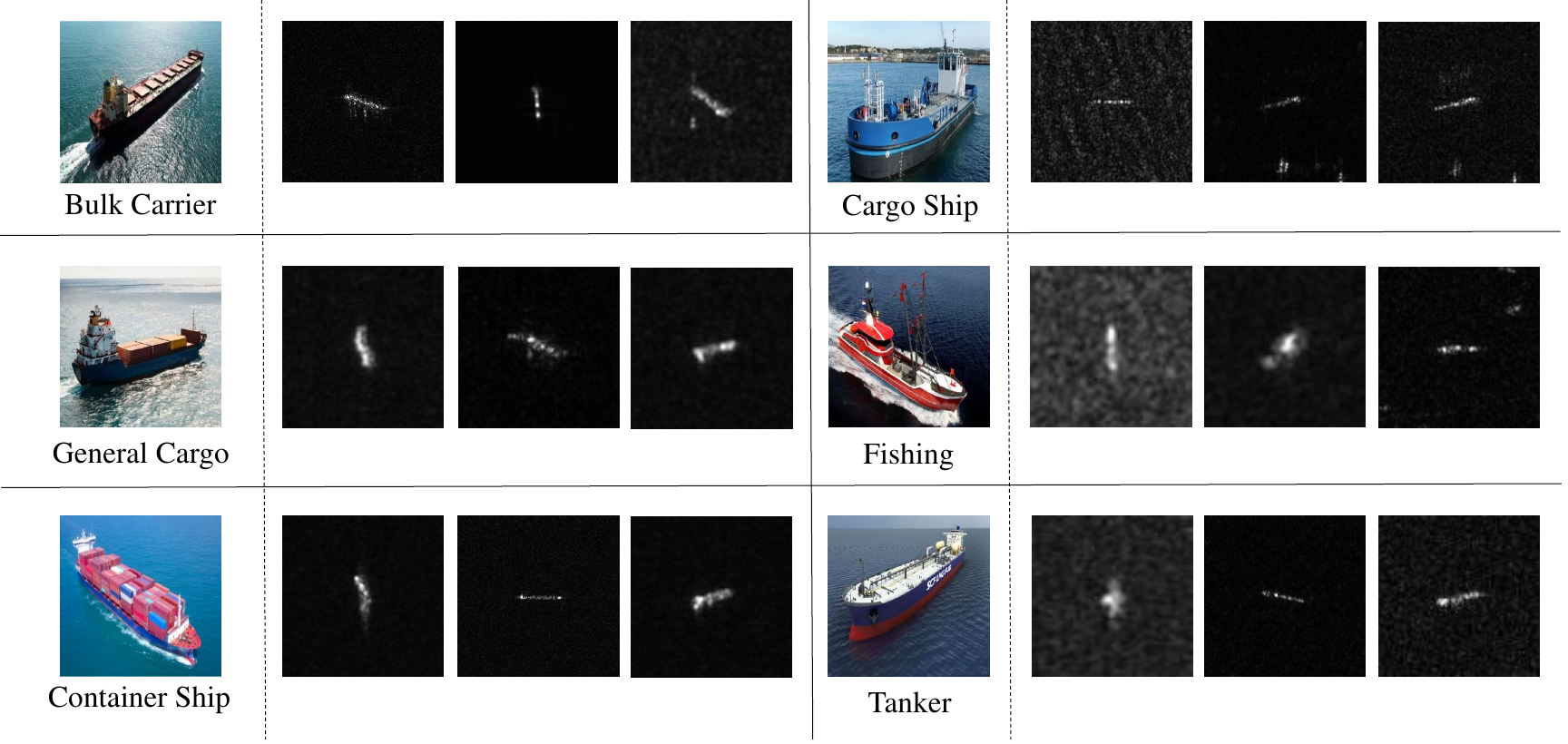} 
\caption{SAR ship images and corresponding optical ship images of six classes in the OpenSARShip dataset.}
\label{sampleOPEN}
\end{figure}

\begin{figure}[thb]
\centering
\includegraphics[width=0.99\linewidth]{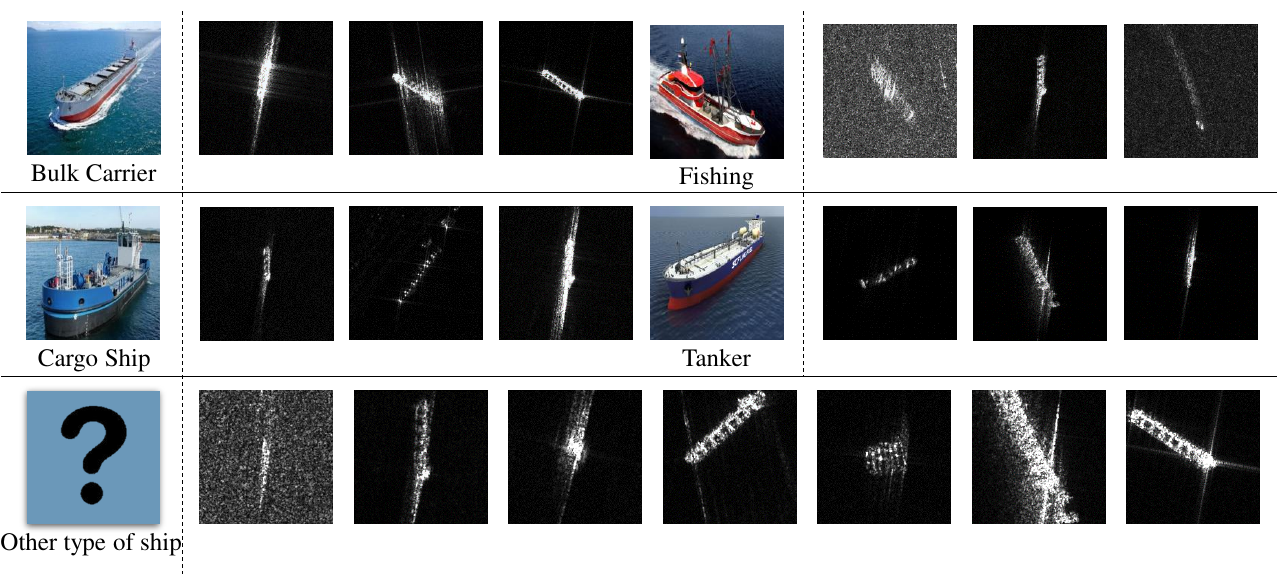} 
\caption{SAR ship images and corresponding optical ship images of six classes in the FUSAR-Ship dataset. The last ship class, other type of ship, is s a gather, which includes most of other ship classes except the common ship classes. This ship class has more overlap with other ship classes and can validate the robustness and effectiveness of our method more comprehensively.}
\label{sampleFUSAR}
\end{figure}

\section{Experiments and Results}
This sections validates the effectiveness and robustness of the proposed method on the OpenSARShip and FUSAR-Ship datasets. To evaluate the proposed method, ablation experiments were conducted, which means the model with/without the main innovation to compare the performance under different ablation configurations.
To evaluate the practical application capabilities, recognition experiments are conducted under decreasing training SAR ship samples. 
The two used benchmark datasets, i.e., OpenSARShip and FUSAR-Ship, and the corresponding preprocessing are first described in detail. The experimental results are then demonstrated for OpenSARShip under six-class and three-class conditions. Moreover, the recognition results on the FUSAR-Ship dataset are provided. 
In the end, the comparison under different training SAR ship samples and the comparison with different effective deep learning networks under constant training SAR ship samples are presented and analyzed.

\subsection{Datasets and Network Configuration}

OpenSARShip and FUSAR-Ship are two benchmark datasets for SAR ship recognition.
The OpenSARShip dataset develops sophisticated ship detection and classification algorithms under high interference. The data were gathered from 41 Sentinel-1 images under disparate environmental conditions. A total of 11,346 ship chips for 17 types of SAR ships were integrated with AIS information. The labels in this dataset are reliable because these labels of ships are based on AIS information \cite{ais}. In this paper, the experiments used the ground range detected (GRD) data, with a resolution of $2.0\textit{m} \times 1.5\textit{m}$ and a pixel size of $10\textit{m} \times 10\textit{m}$ in azimuth and distance directions under Sentinel-1 IW mode. The length and the width of these ships range from 92m to 399m and from 6m to 65m, respectively. Fig. \ref{sampleOPEN} shows six classes of the SAR ship images.

FUSAR-Ship is an open SAR-AIS matchup dataset of the Gaofen-3 (GF-3) satellite compiled by the Key Lab of Information Science of Electromagnetic Waves (Ministry of Education) of Fudan University \cite{FUSAR}. GF-3 is China's first civilian C-band fully polarized satellite-based SAR that is mainly used for marine remote sensing and marine monitoring. The FUSAR-Ship dataset was constructed using an automatic SAR-AIS matchup procedure for more than 100 GF-3 scenes, including over 5000 ship image chips with AIS information. FUSAR-Ship is considered an open benchmark dataset for ship and marine target detection and identification.

The training process and the configuration of the network are described below. By employing bilinear interpolation to the original data, the size of the input SAR image is $224 \time 224$. The ${\lambda_1}$ and ${\lambda_2}$ values are set as 1 and 0.5, respectively. $\psi $ is fixed to 0.1. The feature center $K$ is given a value of 10. $\delta $ is adjusted to 0.1. The batch size is set as 32. The learning rate is initialized to 0.01 and decreases at a rate of 0.5 every 25 epochs. There are also five epochs for training warm-up. The quantity of ${v_1}$, ${v_2}$, ${v_3}$, and ${v_4}$ are set as 2, 2, 8 and 2, respectively. The dropout parameters are modified to 0.5 for training and 1 for testing. The dimension of the first fully connected layer is set as 512. Other parameters are employed, and the settings of the framework and structure are presented in detail, as shown in Figs. \ref{framework} and \ref{structure}.
The proposed method was measured and evaluated on a computer with an Intel Xeon Gold 6130 CPU at 2.10 GHz, two Nvidia TITAN RTX, and eight 32 GB memories. The open-source PyTorch framework was used to implement the proposed method.

\renewcommand{\arraystretch}{1.5}
\begin{table}[]
\centering
\caption{Image Number and Imaging Conditions of Different Targets in OpenSARShip}
\label{opensarset}
\setlength\tabcolsep{1.8pt}
\begin{tabular}{c|c|ccc}
\toprule \toprule
Class          & Imaging Condition                                                                  & \begin{tabular}[c]{@{}c@{}}Training\\ Number\end{tabular} & \begin{tabular}[c]{@{}c@{}}Testing\\ Number\end{tabular} & \begin{tabular}[c]{@{}c@{}}Total\\ Number\end{tabular} \\ \midrule 
Bulk Carrier   & \multirow{6}{*}{\begin{tabular}[c]{@{}c@{}} VH and VV, C ban\\  Resolution $=5-20$m\\ Incident angle $=20^{\circ}-45^{\circ}$ \\ Elevation sweep angle $=\pm 11^{\circ}$\\ ${\text{Rg20}}m \times {\text{az}}22m$\end{tabular}} & 200                                                       & 475                                                      & 675                                                    \\ \cline{1-1} \cline{3-5}
Container Ship &                                                                                    & 200                                                       & 811                                                      & 1011                                                   \\ \cline{1-1} \cline{3-5}
Tanker         &                                                                                    & 200                                                       & 354                                                      & 554                                                    \\ \cline{1-1} \cline{3-5}
Cargo          &                                                                                    & 200                                                       & 557                                                      & 757                                                    \\ \cline{1-1} \cline{3-5}
Fishing        &                                                                                    & 200                                                       & 121                                                      & 321                                                    \\ \cline{1-1} \cline{3-5}
General Cargo  &                                                                                    & 200                                                       & 165                                                      & 365     \\ \bottomrule \bottomrule                                               
\end{tabular}
\end{table}

\begin{figure*}[!htb]
\begin{center}
\subfigure[]{\label{1.1}\includegraphics[width=1.6in]{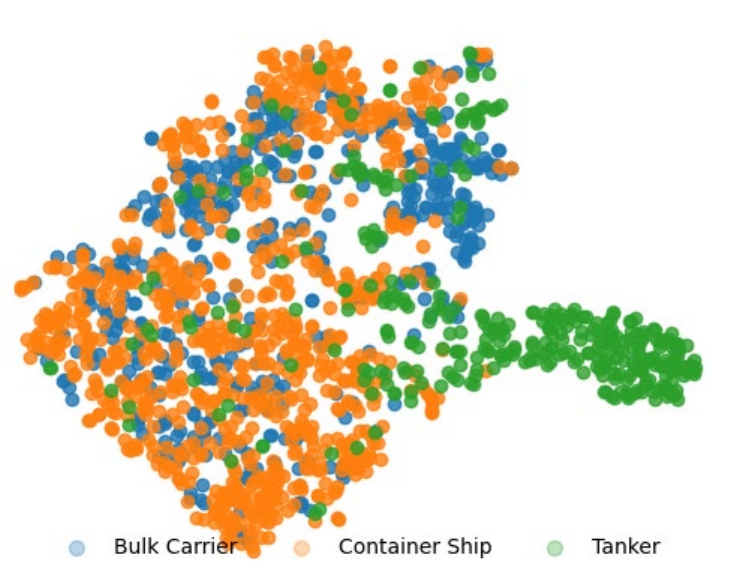}}
\subfigure[]{\label{1.2}\includegraphics[width=1.6in]{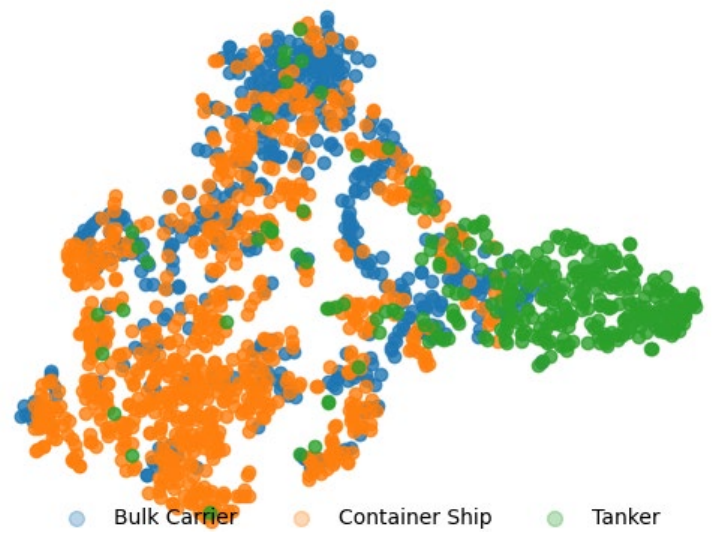}}
\subfigure[]{\label{1.3}\includegraphics[width=1.6in]{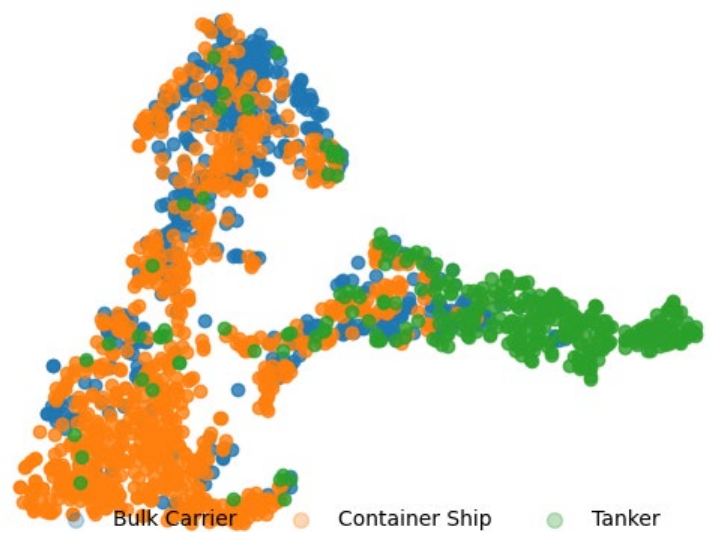}}
\subfigure[]{\label{1.4}\includegraphics[width=1.6in]{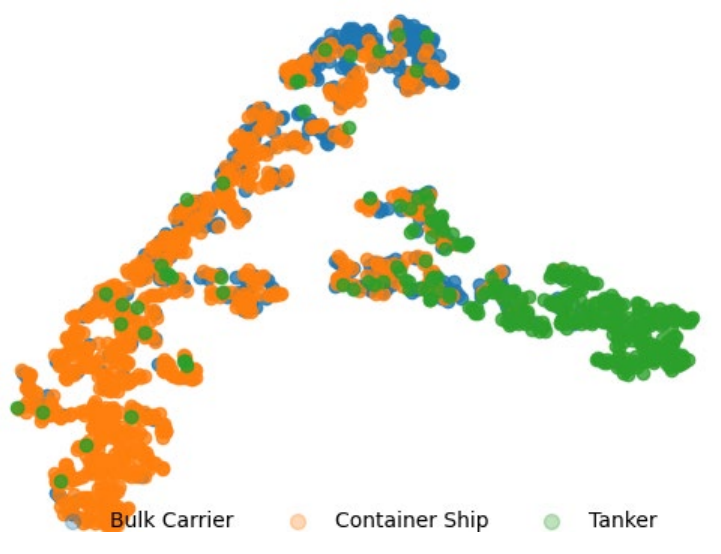}}\\
\end{center}
\caption{Ablation Experiments: Feature Visualization under Different Ablation Configurations. (a) Plain model without the two proposed innovations. (b) Model only with the multifeature center classifier. (c) Model only with the selective feature discrimination. (d) Full version of the model.}
\label{ablation_fig}
\end{figure*}

\renewcommand{\arraystretch}{1.6}
\begin{table}[]
\centering
\small
\caption{Ablation Experiments: Recognition Performance (\%) of Different Ablation Configurations under 40 Training Samples. SFD, Selective Feature Discrimination; MFCC, MultiFeature Center Classifier}
\label{ablation_tab}
\setlength\tabcolsep{4.6pt}
\begin{tabular}{c|cc|cccc}
\toprule \toprule
\multirow{2}{*}{Method} & \multirow{2}{*}{SFD} & \multirow{2}{*}{MFCC} & \multirow{2}{*}{\begin{tabular}[c]{@{}c@{}}Bulk\\Carrier\end{tabular}} & \multirow{2}{*}{\begin{tabular}[c]{@{}c@{}}Container\\Ship\end{tabular}} & \multirow{2}{*}{Tanker} & \multirow{2}{*}{Average} \\
 &  &  &  &  &  &  \\ \midrule 
V1 & × & × & 54.95 & 61.41 & 79.66 & 63.48 \\ \cline{1-1} \cline{4-7}
V2 & \checkmark & × & 57.05 & 68.43 & 87.29 & 69.21 \\ \cline{1-1} \cline{4-7}
V3 & × & \checkmark & 58.32 & 69.42 & 88.70 & 70.37 \\ \cline{1-1} \cline{4-7}
Ours & \checkmark & \checkmark & 82.74 & 73.98 & 77.97 & 77.42 \\ \bottomrule \bottomrule
\end{tabular} 
\end{table}

\renewcommand{\arraystretch}{1.6}
\begin{table*}[]
\centering
\caption{Ablation Experiments with Other Metric Learning-based Classifiers in OpenSARShip}
\label{metric_comparion}
\begin{tabular}{c|c|ccccccccc}
\toprule \toprule 
\multirow{2}{*}{} & \multirow{2}{*}{Methods} & \multicolumn{9}{c}{Training Number in Each Class} \\ \cline{3-11}
 &  & 10 & 20 & 30 & 40 & 60 & 70 & 80 & 100 & 200 \\ \midrule 
\multirow{4}{*}{6-class} & Arcface \cite{arcface} & 42.77\% & 45.79\% & 46.23\% & 49.98\% & 49.78\% & 52.88\% & 55.98\% & 56.46\% & 60.25\% \\ \cline{2-11}
 & Triplet \cite{triplet} & 47.08\% & 49.09\% & 53.69\% & 54.45\% & 54.53\% & 55.38\% & 56.95\% & 61.98\% & 64.48\% \\ \cline{2-11}
 & Sphere \cite{sphere} & 47.81\% & 51.11\% & 51.71\% & 55.62\% & 59.00\% & 59.36\% & 60.73\% & 57.43\% & 66.13\% \\ \cline{2-11}
 & Ours  &48.38\% & 53.41\% & 54.32\% & 64.53\% & 63.88\% & 70.11\% & 68.69\% & 70.15\% & 76.85\%  \\  \midrule 
\multirow{4}{*}{3-class} & Arcface \cite{arcface} & 63.54\% & 69.70\% & 70.43\% & 71.04\% & 70.24\% & 73.23\% & 74.39\% & 73.29\% & 77.32\% \\ \cline{2-11}
 & Triplet \cite{triplet} & 65.24\% & 69.09\% & 70.30\% & 70.73\% & 72.68\% & 76.83\% & 77.50\% & 78.35\% & 79.63\% \\ \cline{2-11}
 & Sphere \cite{sphere} & 64.45\% & 67.50\% & 69.57\% & 70.24\% & 72.99\% & 76.89\% & 78.17\% & 79.70\% & 80.43\% \\ \cline{2-11}
 & Ours & 68.85\% & 72.15\% & 73.40\% & 77.42\% & 79.13\% & 80.68\% & 81.50\% & 83.47\% & 86.04\% \\
 \bottomrule \bottomrule 
\end{tabular}
\end{table*}

\renewcommand{\arraystretch}{1.8}
\begin{table*}[]
\centering
\caption{Recognition Performance of Three Classes under Different Training Data in OpenSARShip}
\label{open3result}
\begin{tabular}{c|ccccccccc}
\toprule \toprule 
\multirow{2}{*}{Class} & \multicolumn{9}{c}{Training Number in Each Class}                              \\ \cline{2-10} 
                       & 10      & 20      & 30      & 40      & 60      & 70      & 80      & 100     & 200     \\ \midrule 
Bulk Carrier           & 72.63\% & 77.68\% & 79.58\% & 82.74\% & 86.74\% & 86.95\% & 89.68\% & 89.89\% & 92.84\% \\
Container Ship         & 68.06\% & 70.04\% & 71.52\% & 73.98\% & 73.86\% & 75.46\% & 75.46\% & 78.42\% & 81.50\% \\
Tanker                 & 64.97\% & 69.21\% & 68.93\% & 77.97\% & 80.79\% & 83.90\% & 83.90\% & 85.88\% & 87.01\% \\ \midrule 
Average                & 68.85\% & 72.15\% & 73.40\% & 77.42\% & 79.13\% & 80.68\% & 81.50\% & 83.47\% & 86.04\% \\ \bottomrule \bottomrule 
\end{tabular} 
\end{table*}

\renewcommand{\arraystretch}{1.8}
\begin{table*}[]
\centering
\caption{Recognition Performance of Six Classes under Different Training Data in OpenSARShip}
\label{open6result}
\begin{tabular}{c|ccccccccc}
\toprule \toprule
\multirow{2}{*}{Class} & \multicolumn{9}{c}{Training Number in Each Class}                              \\ \cline{2-10} 
                       & 10      & 20      & 30      & 40      & 60      & 70       & 80      & 100     & 200     \\ \midrule 
Bulk Carrier           & 53.26\% & 58.74\% & 51.58\% & 66.11\% & 66.11\% & 69.05\% & 66.11\% & 70.11\% & 70.74\% \\
Container Ship         & 64.12\% & 71.52\% & 68.80\% & 77.07\% & 75.09\% & 82.00\% & 81.01\% & 85.08\% & 92.11\% \\
Tanker                 & 40.11\% & 37.29\% & 47.46\% & 53.11\% & 51.69\% & 59.32\% & 57.91\% & 55.93\% & 72.03\% \\
Cargo                  & 27.11\% & 31.78\% & 36.98\% & 48.65\% & 48.83\% & 58.89\% & 59.07\% & 57.99\% & 64.99\% \\
Fishing                & 88.43\% & 77.69\% & 88.43\% & 90.91\% & 95.04\% & 95.04\% & 90.08\% & 91.74\% & 90.08\% \\
General Cargo          & 15.76\% & 37.58\% & 37.58\% & 56.36\% & 54.55\% & 55.76\% & 54.55\% & 51.52\% & 58.18\% \\ \midrule 
Average                & 48.38\% & 53.41\% & 54.32\% & 64.53\% & 63.88\% & 70.11\% & 68.69\% & 70.15\% & 76.85\% \\ \bottomrule \bottomrule
\end{tabular} 
\end{table*}

\subsection{Method Soundness Verification}

This section evaluates the soundness verification experiments of the proposed method by running ablation experiments. Two types of ablation experiments were conducted. The first type is recognition experiments of different methods with/without the partial feature discrimination and multifeature center classifier. The second type is recognition experiments of the methods with the partial feature discrimination, but with different metric learning-based classifiers. The second type of ablation experiments aims to compare the effectiveness of our multifeature center classifier.

\subsubsection{Ablation Experiments under Different Configurations}

There are four comparison experiments with 40 training samples of each class under different ablation configurations. The first configuration, V1, is the plain network without the two main innovations, i.e., selective feature discrimination and multifeature center classifier. The second one, V2, is the method with the proposed selective feature discrimination but without the multifeature center classifier. The third one, V3, is the method with the proposed multifeature center classifier but without the selective feature discrimination. The last one, Ours, is the full version of our method.

Following \cite{open2,open3,open4,open5}, three classes of ships, i.e., bulk carrier, container ship, and tanks, are chosen from OpenSARShip. These three classes of ships are the most common and representative ships occupying 80\% of the international shipping market \cite{open5}.
Table \ref{opensarset} presents the original dataset of the training and testing sets.

It is uneven for the number of objects in the three classes. To avoid the effect of class imbalance, the training and testing ratio was set to 4:6 based on the smallest number of samples among the three classes of objects, and the number of training samples in each class was made equal.
Under decreasing limited training data, different numbers of the training data from the 200 original training sets were randomly chosen to validate the proposed method. The different training numbers were augmented to 200 for each class by random sampling to reduce the effect of the different training data numbers \cite{augment}. Before training, 40 samples of each class were randomly chosen from the original 200 samples.

Table \ref{ablation_tab} and Fig. \ref{ablation_fig} show the recognition results and feature visualization, respectively.
As presented in Table \ref{ablation_tab}, the recognition ratio of the first configuration (without the two innovations) is 63.48\%. The recognition ratio of the methods in this paper is 77.42\%. It is clear that the innovations can clearly improve the recognition performance.
As shown in Fig. \ref{ablation_fig}, it is clear that the plain model in (a) cannot handle the large inner-class variance and small interclass differences and fails with low recognition ratios. 
By comparing the four subfigures, it is clear that the two innovations can boost the confusing feature distributions of the plain model shown in Fig. \ref{ablation_fig} (a) into the more discriminative feature distributions of the model shown in Fig. \ref{ablation_fig} (d). Furthermore, the gradual improvements of the feature distributions from Fig. \ref{ablation_fig} (a)-(d) correspond to the improvements presented in Table \ref{ablation_tab}.

\subsubsection{Ablation Experiments Using different metric learning-based classifiers}

In this recognition experiment, the base method is the vanilla model with the proposed selective feature discrimination but without the multifeature center classifier, V2, as presented in Table \ref{ablation_tab}. Three effective metric learning-based classifiers, i.e., arcface \cite{arcface}, triplet \cite{triplet}, and sphere \cite{sphere}, are then chosen to be compared with the proposed multifeature center classifier. Arcface \cite{arcface} proposes a geodesic distance on a hypersphere as an additive angular margin loss to obtain highly discriminative features. 
Triplet \cite{triplet} is a famous metric learning-based classifier, which consist of an anchor sample, a positive sample of the same class as the anchor, and a negative sample of a different class. The goal is to learn a feature embedding such that the distance between the anchor and the positive sample is smaller than the distance between the anchor and the negative sample by a margin.
Sphere \cite{sphere} proposes an angular SoftMax loss that enables convolutional neural networks (CNNs) to learn discriminative features.
Table \ref{metric_comparion} presents the comparison of recognition performances of the three metric learning-based classifiers and the proposed multifeature center classifier. 
It is clear that the proposed multifeature center classifier can achieve superior recognition ratios when the training samples range from 10 to 200 for each class. 

From the comparison of two ablation experiments in recognition ratios and feature visualizations, the recognition ratios of the proposed method obviously outperform the plain model without the proposed method or just with one of the two innovations. The feature visualizations of four different ablation configurations also show the effectiveness of the proposed method and the reason of the higher recognition ratios.
Therefore, the soundness of our method has been validated.

\subsection{Recognition Results of the three and six classes of ships under OpenSARShip}
This section selects two types of recognition experiments, i.e., those with three classes and six classes of ships, to comprehensively validate the proposed method. Based on the three classes mentioned above, another three classes, i.e., cargo ship, fishing, and general cargo, were chosen to organize one more challenging six-class recognition experiment.

Table \ref{open3result} presents the recognition results of the proposed method under three classes of training data in OpenSARShip. It is clear that when there are 200 training data for each class, the overall recognition ratio can achieve 86.04\%. At the same time, when the training data range from 100 to 60 for each ship class, the recognition ratios decrease from 83.47\% to 79.13\%, denoting that the proposed method is robust and effective in this range. Facing the decreasing training set from 30 to 20, the recognition ratios are stable at around 73.50\%, which is an inspiring performance of the SAR ship target recognition. Under the rough limited training data, i.e., ten for each class, the recognition ratio of SAR ship target recognition is still 68.85\%.

Table \ref{open6result} presents the recognition results of the proposed method under six classes of training data in OpenSARShip. The six-class recognition is more challenging due to the large similarity of the different ship classes and large variance of the same ship classes. The SAR ship target recognition is 76.85\% for 200 training samples for each class. With the limited training data decreasing from 100 to 60, the recognition ratios of the proposed method are also stable, decreasing from 70.15\% to 63.88\%. This six-class performance of the SAR ship target recognition is similar to the three-class performance in terms of robustness. When the training sample continues to decrease from 40 to 20, the recognition ratios are still stable with small variance, from 64.53\% to 53.41\%. Facing ten training samples for each class, the recognition ratio of the proposed method achieves 48.38\%. Such situations are frequently the actual ones that will be faced, which are crucial challenges for practical applications.

From the quantitative results, it was shown that the proposed method not only achieves high performance of SAR ship target recognition but also shows the robustness and effectiveness under decreasing limited training data. To further validate the proposed method, the experiments under FUSAR-Ship dataset are shown as follows.

\subsection{Recognition Results of five and three classes of ships under FUSAR-Ship}

\renewcommand{\arraystretch}{1.5}
\begin{table}[]
\centering
\caption{Image Number and Imaging Conditions of Different Targets in FUSAR-Ship}
\label{fusarset}
\setlength\tabcolsep{0.5pt}
\begin{tabular}{c|c|ccc}
\toprule \toprule 
Class          & Imaging Condition                                                                  & \begin{tabular}[c]{@{}c@{}}Training\\ Number\end{tabular} & \begin{tabular}[c]{@{}c@{}}Testing\\ Number\end{tabular} & \begin{tabular}[c]{@{}c@{}}Total\\ Number\end{tabular} \\ \midrule  
Bulk Carrier          & \multirow{5}{*}{\begin{tabular}[c]{@{}c@{}}VH and VV, C band\\  Resolution $=0.5-500$m\\ Incident angle $=10^{\circ}-60^{\circ}$ \\ Elevation sweep angle $=\pm 20^{\circ}$\\ ${\text{Rg20}}m \times {\text{az}}22m$\end{tabular}} & 100                  & 173                  & 273                  \\ \cline{1-1} \cline{3-5}
Cargo Ship            &                                                                                    & 100                  & 1593                 & 1693                 \\ \cline{1-1} \cline{3-5}
Fishing              &                                                                                    & 100                  & 685                  & 785                  \\ \cline{1-1} \cline{3-5}
Other type of ship      &                                                                                    & 100                  & 1507                 & 1607                 \\ \cline{1-1} \cline{3-5}
Tanker               &                                                                                    & 100                  & 48                   & 148                  \\ \bottomrule \bottomrule                                               
\end{tabular} 
\end{table}

Though there are several dozens of ship classes in the FUSAR-Ship dataset, the numbers of the SAR ship images of most ship classes are less than 70. Therefore, limited by the number distributions of all the ship classes in the FUSAR-Ship dataset, five classes, i.e., bulk carrier, cargo ship, fishing, tanker, and other types of ship, which have more than 100 SAR ship images were chosen. Bulk carrier, cargo, fishing, and tanker ships are the most common ship classes in the world shipping market. The last ship class, i.e., other types of ship, is a gather , which includes most of other ship classes except the common ship classes. This ship class has more overlap with other ship classes and can validate the robustness and effectiveness of the method more comprehensively. Table \ref{fusarset} presents the original dataset of the training and testing sets. 

As with the preprocessing of the OpenSARShip dataset, an equal number of training samples were also set for each class based on a minimum number of training and testing ratios of 4:6 for all three classes of samples . For a decreasing limited training data, different numbers of the training data were randomly chosen from the 100 original training sets to validate the proposed method. The different training numbers were also augmented to 100 for each class through random sampling.

\renewcommand{\arraystretch}{1.5}
\begin{table*}[]\small
\centering
\caption{Recognition Performance of Five Classes under Different Training Data in FUSAR-Ship}
\label{fusar5results}
\begin{tabular}{c|cccccc}
\toprule \toprule
\multirow{2}{*}{Class} & \multicolumn{6}{c}{Training Number in Each Class}          \\ \cline{2-7} 
                       & 20      & 30      & 40      & 60      & 80      & 100     \\ \midrule 
Bulk carrier            & 83.40\% & 80.25\% & 85.45\% & 88.84\% & 52.79\% & 82.66\% \\
Cargo ship              & 49.97\% & 57.37\% & 53.95\% & 60.25\% & 78.00\% & 62.21\% \\
Fishing                & 43.40\% & 39.60\% & 69.10\% & 52.35\% & 47.98\% & 61.17\% \\
Other type of ship     & 63.01\% & 68.23\% & 73.76\% & 74.86\% & 96.40\% & 82.08\% \\
Tanker                 & 78.13\% & 59.32\% & 79.55\% & 78.70\% & 18.73\% & 85.42\% \\ \midrule 
Average                & 56.26\% & 59.55\% & 65.98\% & 66.21\% & 70.38\% & 70.67\% \\ \bottomrule \bottomrule 
\end{tabular} 
\end{table*}

\renewcommand{\arraystretch}{1.5}
\begin{table*}[]\small
\centering
\caption{Recognition Performance of Three Classes under Different Training Data in FUSAR-Ship}
\label{fusar3results}
\begin{tabular}{c|cccccc}
\toprule \toprule
\multirow{2}{*}{Class} & \multicolumn{6}{c}{Training Number in Each Class}          \\ \cline{2-7} 
                       & 20      & 30      & 40      & 60      & 80      & 100     \\ \midrule 
Bulk carrier            & 86.56\% & 78.60\% & 84.55\% & 78.40\% & 80.83\% & 80.35\% \\
Cargo ship              & 51.64\% & 65.54\% & 59.35\% & 63.93\% & 66.77\% & 67.61\% \\
Fishing                & 62.22\% & 45.43\% & 59.19\% & 58.21\% & 60.14\% & 63.50\% \\ \midrule 
Average                & 57.93\% & 61.03\% & 61.54\% & 63.52\% & 65.99\% & 67.36\% \\ \bottomrule \bottomrule 
\end{tabular} 
\end{table*}

Table \ref{fusar5results} presents the recognition results of the proposed method under five classes of training data in the FUSAR-Ship dataset. The proposed method achieves a recognition result of 70.66\% under 100 training images for each ship class. When the training data decreases from 100 to 80, the recognition ratio of the proposed method decreases from 70.66\% to 70.38\%. Facing 60 training images for each ship class, the overall recognition ratio of the proposed method still achieves 66.20\%. When the training data is reduced from 60 to 40, the proposed method can achieve a recognition ratio of 65.97\%. When the training data range from 30 to 20, the recognition ratios of the proposed method range from 59.55\% to 56.26\%, indicating that the proposed method is robust for the decreasing training data.

Table \ref{fusar3results} presents the recognition results of the proposed method under three classes of training data in the FUSAR-Ship dataset. It is clear that the proposed method shows robust and effective recognition performance when the training samples for each class are 100 and 80. The corresponding recognition ratios of the proposed method are 67.36\% and 65.98\% under 100 and 80 ship images for each class, respectively. When the training samples for each class are decreased to 60, the recognition ratio of the proposed method is 63.51\%. If the training samples for each class are 40 and 30, the recognition ratios are 61.53\% and 61.02\%, respectively. When the training data continues to be even reduced to 20 for each class, the recognition ratio is 57.93\%, indicating that the proposed method is robust for the decreased training samples. 
From the quantitative experimental results above, it is clear that the proposed method is robust and effective under the decreasing training data in the FUSAR-Ship dataset.

Looking at the results of OpenSARShip and FUSAR-Ship datasets, the proposed method proves to be effective and robust under the experiments with limited training samples. Furthermore, the proposed method can achieve excellent recognition performance in facing different SAR datasets. The comparison with other methods will then be presented as follows.

\section{Conclusion}
SAR ship target recognition is a crucial and basic research field among the application of SAR ATR for ship monitoring. 
Due to the large inner-class variance and small interclass differences of SAR ship images, the existing method inputs all the extracted features of SAR ship images directly into one classifier to complete the recognition. Moreover, the classifier of most methods is with a single feature center for each class. 
Therefore, the performances of these methods are limited. 
The recognition method in this paper proposed a novel framework to gradually extract the features, enhance the useless partial features with more discriminativeness to tackle small interclass differences, and assign robust multifeature centers for large inner-class variance. 
Finally, the method achieved accurate recognition performance of SAR ship images. 
From the experimental results on the OpenSARShip and FUSAR-Ship datasets, the proposed methods  have greatly improved the recognition performance under decreasing training samples. It was also illustrated that the method is robust and effective for the experiments with increasing ship class numbers. From the comparison with other methods of SAR ship recognition and widely used deep learning networks, it is clear that the proposed method has achieved the state-of-the-art performance.

\bibliographystyle{IEEEtran}
\bibliography{references}

\newpage

\begin{IEEEbiography}[{\includegraphics[width=1in,height=1.25in,clip,keepaspectratio]{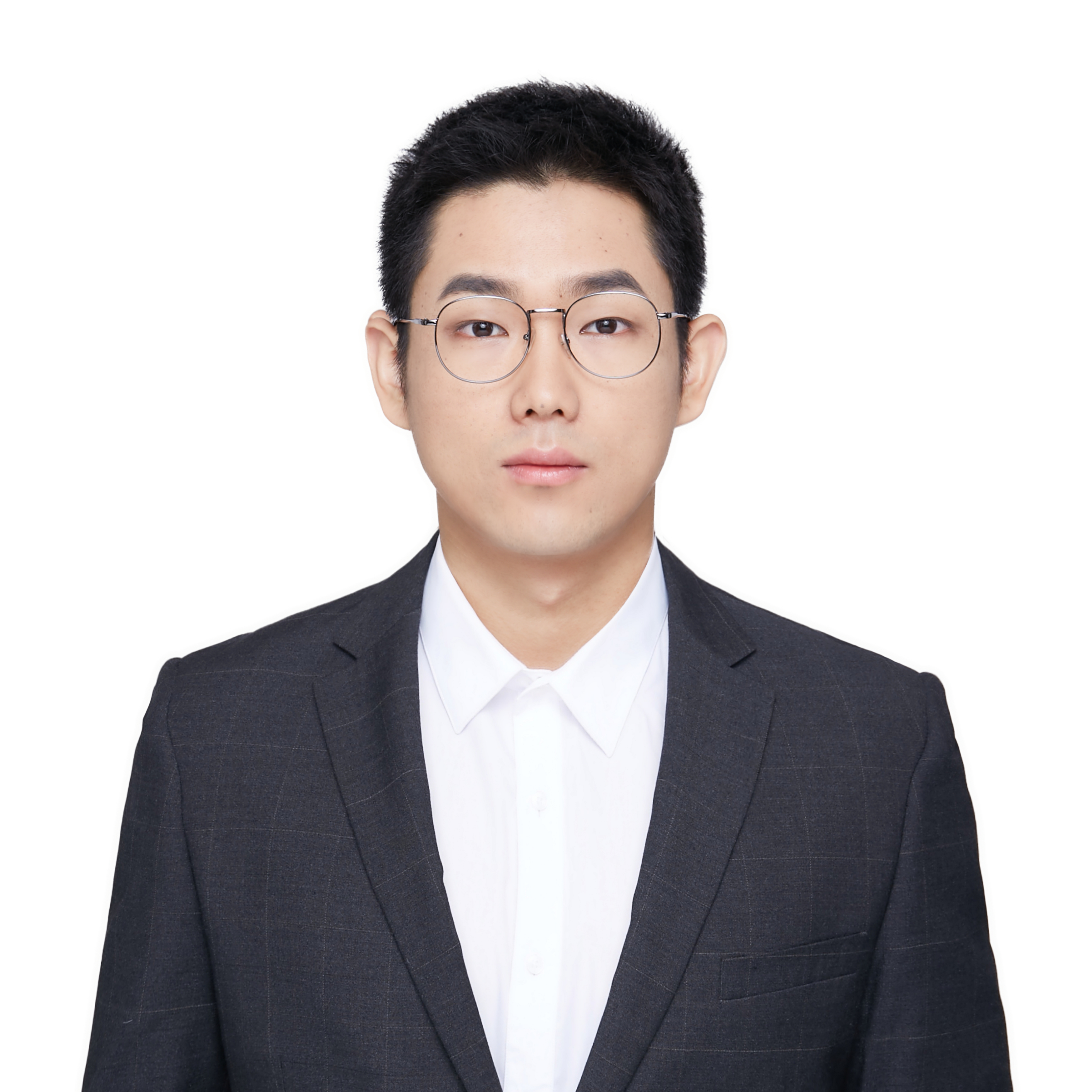}}]{Chenwei Wang}
received the B.S. degree from the School of Electronic Engineering, University of Electronic Science and Technology of China (UESTC), Chengdu, China, in 2018. He is currently pursuing the Ph.D. degree with the School of Information and Communication Engineering, University of Electronic Science and Technology of China, Chengdu, China.
His research interests include radar signal processing, machine learning, and automatic target recognition.
\end{IEEEbiography}

\begin{IEEEbiography}[{\includegraphics[width=1in,height=1.25in,clip,keepaspectratio]{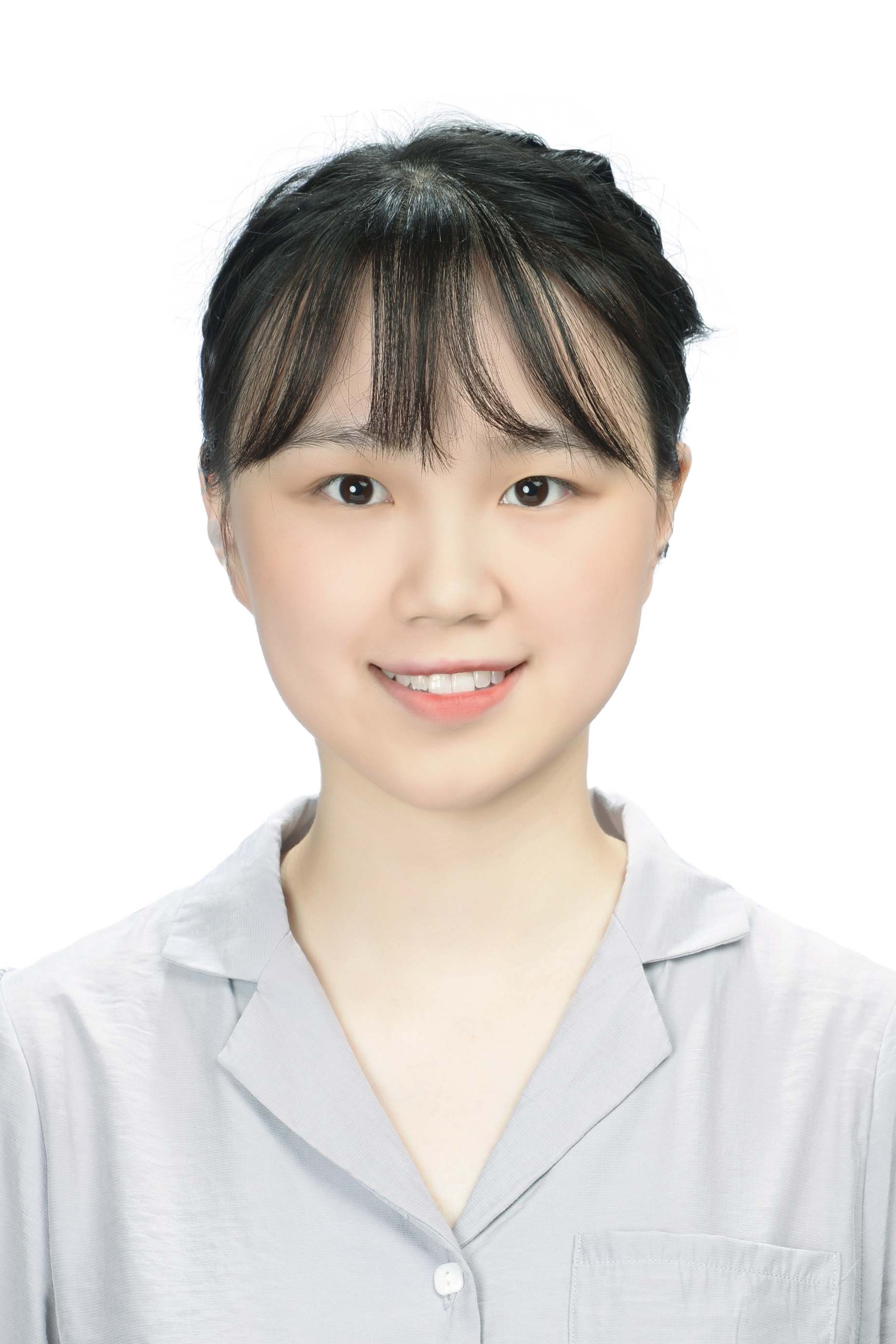}}]{Siyi Luo}
received the B.S. dual degree from Chongqing University (CQU), Chongqing, China, in 2021. She is currently pursuing the M.S. degree with the School of Information and Communication Engineering, University of Electronic Science and Technology of China, Chengdu, China. Her research interests include machine learning, target detection, and automatic target recognition.
\end{IEEEbiography}

\begin{IEEEbiography}[{\includegraphics[width=1in,height=1.25in,clip,keepaspectratio]{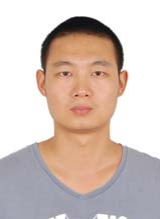}}]{Jifang Pei}
(M'19) received the B.S. degree from the College of Information Engineering, Xiangtan University, Hunan, China, in 2010, and the M.S. degree from the School of Electronic Engineering, University of Electronic Science and Technology of China (UESTC), Chengdu, China, in 2013. He received the Ph.D. degree from the School of Information and Communication Engineering, UESTC, in 2018. From 2016 to 2017, he was a joint Ph.D. Student with the Department of Electrical and Computer Engineering, National University of Singapore, Singapore. He is currently an Associate Research Fellow with the School of Information and Communication Engineering, UESTC. His research interests include radar signal processing, machine learning, and automatic target recognition.
\end{IEEEbiography}

\begin{IEEEbiography}[{\includegraphics[width=1in,height=1.25in,clip,keepaspectratio]{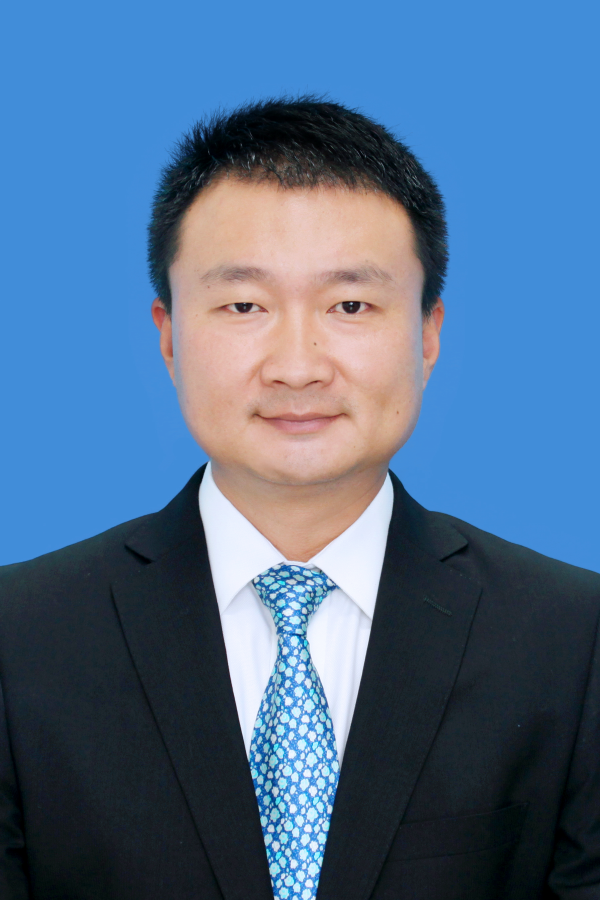}}]{Yulin Huang}
(M'08) received the B.S. and Ph.D. degrees from the School of Electronic Engineering, University of Electronic Science and Technology of China (UESTC), Chengdu, China, in 2002 and 2008, respectively. He is currently a Professor at the UESTC. His research interests include radar signal processing and SAR automatic target recognition.
\end{IEEEbiography}

\begin{IEEEbiography}[{\includegraphics[width=1in,height=1.25in,clip,keepaspectratio]{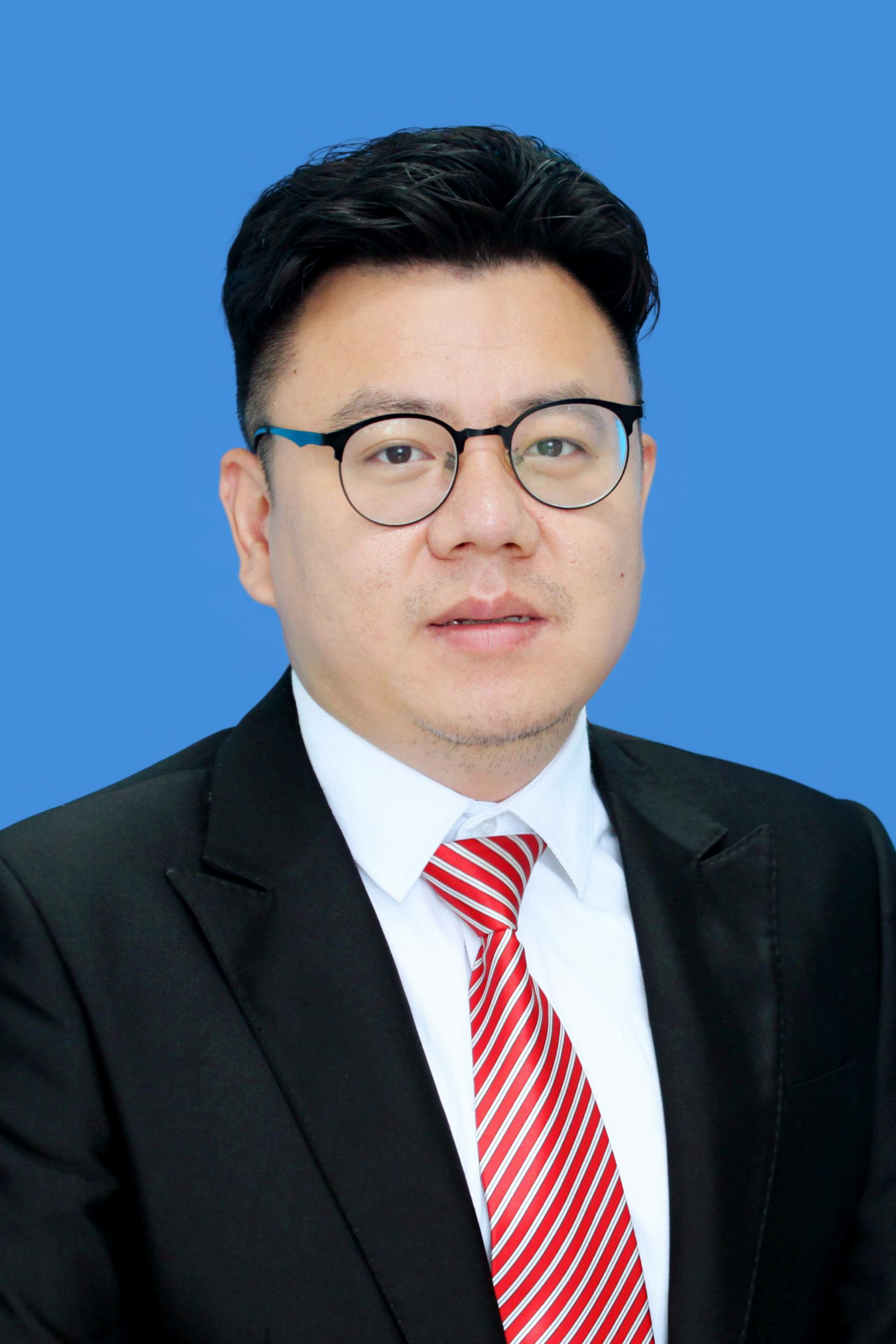}}]{Yin Zhang}
(M'16) received the B.S. and Ph.D. degrees from the School of Electronic Engineering, University of Electronic Science and Technology of China (UESTC), Chengdu, China, in 2008 and 2016, respectively. From September 2014 to September 2015, he had been a Visiting Student with the Department of Electrical and Computer Engineering, University of Delaware, Newark, USA. He is currently an Research Fellow at the UESTC. His research interests include signal processing and radar imaging.
\end{IEEEbiography}

\begin{IEEEbiography}[{\includegraphics[width=1in,height=1.25in,clip,keepaspectratio]{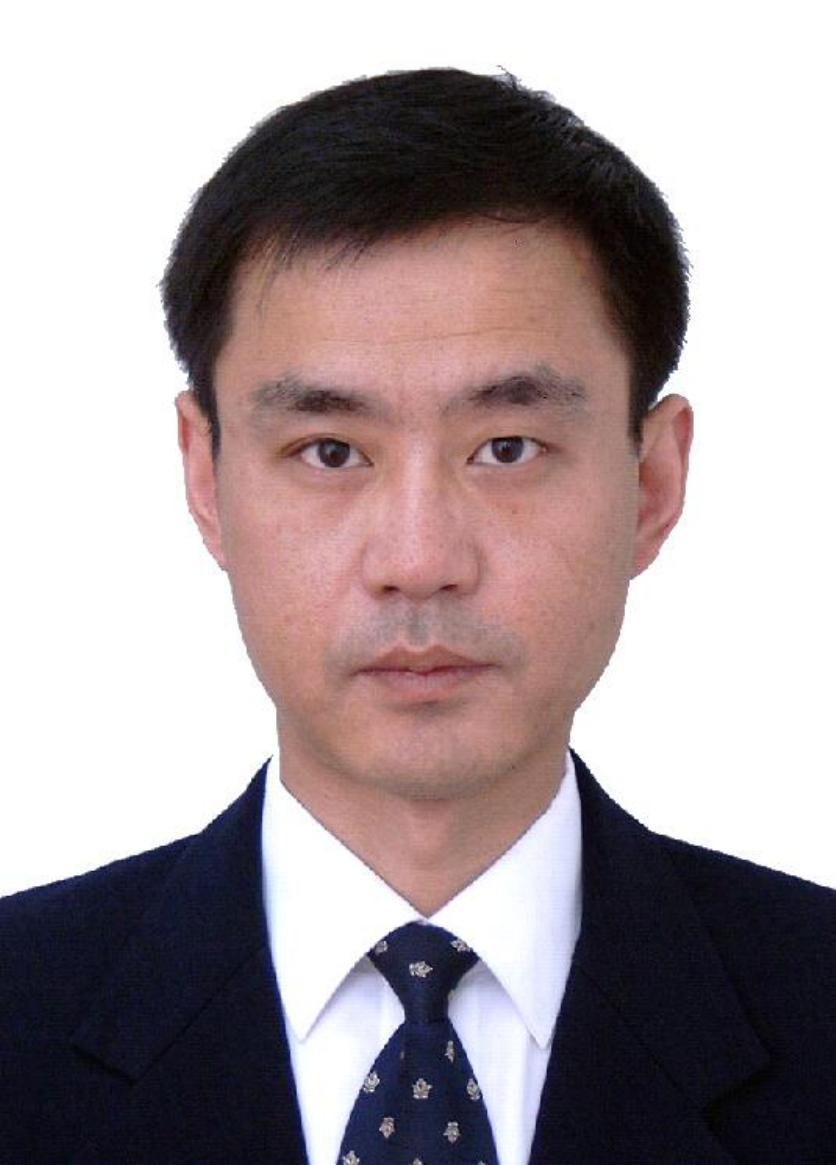}}]{Jianyu Yang}
(M'06) received the B.S. degree from the National University of Defense Technology, Changsha, China, in 1984, and the M.S. and Ph.D. degrees from the University of Electronic Science and Technology of China (UESTC), Chengdu, China, in 1987 and 1991, respectively. He is currently a Professor with the UESTC. From 2001 to 2005, he served as the Dean of School of Electronic Engineering of UESTC. In 2005, he was a Senior Visiting Scholar with the Massachusetts Institute of Technology (MIT), Cambridge, MA, USA. He was selected as the Vice-Chairman of the Radar Society of the Chinese Institute of Electronics (CIE) in 2016 and a Fellow of CIE in 2018. He serves as a Senior Editor for the Chinese Journal of Radio Science and the Journal of Systems Engineering and Electronics. 
His research interests include synthetic aperture radar imaging and automatic target recognition.
\end{IEEEbiography}


%






\end{document}